\documentclass{article}%
\usepackage{amsfonts}
\usepackage{amsmath}
\usepackage{amssymb}
\usepackage{graphicx}%
\providecommand{\U}[1]{\protect\rule{.1in}{.1in}}

\begin{document}

\title{Quantum phases in entropic dynamics\thanks{Presented at MaxEnt 2017, the 37th
International Workshop on Bayesian Inference and Maximum Entropy Methods in
Science and Engineering (July 9-14, 2017, Jarinu, Brazil). }}
\author{Nicholas Carrara and Ariel Caticha\\{\small Department of Physics, University at Albany--SUNY, Albany, NY 12222,
USA}}
\date{}
\maketitle

\begin{abstract}
In the Entropic Dynamics framework the dynamics is driven by maximizing
entropy subject to appropriate constraints. In this work we bring Entropic
Dynamics one step closer to full equivalence with quantum theory by
identifying constraints that lead to wave functions that remain single-valued
even for multi-valued phases by recognizing the intimate relation between
quantum phases, gauge symmetry, and charge quantization.

\end{abstract}

\section{Introduction}

In the Entropic Dynamics (ED) framework the Schr\"{o}dinger equation is
derived as an application of entropic methods of inference\footnote{The
principle of maximum entropy as a method for inference can be traced to the
pioneering work of E. T. Jaynes. For a pedagogical overview of Bayesian and
entropic inference and further references see \cite{Caticha 2012}.} and, as
always with inference, the first and most crucial step is to be clear about
what we want to infer. What microstates are we talking about? This defines the
ontology of the model. Once that choice is made the dynamics is driven by
entropy subject to information expressed by constraints \cite{Caticha
2010a}-\cite{Caticha 2015}.

ED takes the epistemic view of the wave function $\Psi$ to its logical
conclusion. Within an inferential framework it is not sufficient to just state
that the probability $|\Psi|^{2}$ reflects a state of knowledge; it is also
necessary to demand that the phase receive an epistemic interpretation, and
that all changes in $\Psi$ be dictated by the maximum entropy and Bayesian
updating rules. Thus, the ED framework is very restrictive: it must account
for \emph{both} the unitary time evolution described by the Schr\"{o}dinger
equation \emph{and} the collapse of the wave function during measurement.

But even after ED succeeds in accomplishing these tasks a challenge still
remains: is ED fully equivalent to quantum mechanics (QM) or does it merely
reproduce a subset of its solutions? Problems of this kind were pointed out
long ago by Takabayasi \cite{Takabayasi 1952} in the context of the
hydrodynamical interpretation of QM, and later revived by Wallstrom
\cite{Wallstrom 1989}\cite{Wallstrom 1994} in the context of Nelson's
stochastic mechanics. Wallstrom's objection is that stochastic mechanics leads
to phases and wave functions that are either both multi-valued or both
single-valued. Both alternatives are unsatisfactory because on one hand QM
requires single-valued wave functions, while on the other hand single-valued
phases exclude states that are physically relevant (\emph{e.g.}, states with
non-zero angular momentum).

In previous work the constraints that drive the dynamics where introduced in
two different ways, either by postulating some extra variables \cite{Caticha
2010a}\cite{Caticha 2017}, or by the explicit introduction of a
\textquotedblleft drift\textquotedblright\ potential \cite{Caticha
2014a}-\cite{Bartolomeo Caticha 2015}. One of the goals of this paper is to
show that these two types of constraint can be imposed simultaneously which
lends the theory greater flexibility and expands the range of future potential
applications. We identify constraints that lead to single-valued wave
functions, but nevertheless allow for multi-valued phases\footnote{A hint
towards a satisfactory resolution of Wallstrom's objection is found in
Takabayasi's later work which incorporates spin into his hydrodynamical
approach \cite{Takabayasi 1983}. Although here we focus on non-spinning
particles our choice of constraints can be generalized to particles with spin
$1/2$ --- a project to be addressed in a future publication.} and naturally
lead to the local gauge symmetry required for electromagnetic interactions.
Our argument involves two ingredients. The first is the recognition that a
deeper understanding of the phase of the wave function must consider the
intimate relation between quantum phases and gauge symmetry. The second
ingredient is the recognition that in order for ED to agree with experiment it
is necessary that the dynamics be linear. ED differs from standard QM\ in many
crucial ways but its demand for linearity is not one of them. The demand that
the linear and the probabilistic structures be compatible with each other
implies that ED constraints must lead to single-valued wave functions
\cite{Merzbacher 1962}.

Here we will focus on the derivation of the Schr\"{o}dinger equation but the
ED approach has been applied to a variety of other topics including the
quantum measurement problem \cite{Johnson Caticha 2011}\cite{Vanslette Caticha
2016}; momentum and uncertainty relations \cite{Nawaz Caticha 2011}; the
Bohmian limit \cite{Bartolomeo Caticha 2015}\cite{Bartolomeo Caticha 2016} and
the classical limit \cite{Demme Caticha 2016}; the extensions to curved spaces
\cite{Nawaz et al 2015} and to relativistic fields \cite{Ipek Caticha
2014}\cite{Ipek Abedi Caticha 2016}.

\section{Entropic Dynamics --- a brief overview}

\textbf{The statistical model--} We consider $N$ particles living in a flat
Euclidean space $\mathbf{X}$ with metric $\delta_{ab}$. The first important
assumption is that position plays a distinguished role: it defines the ontic
state of the system. The fact that at all times particles have \emph{definite}
positions deviates from the standard Copenhagen interpretation according to
which definite values are created by measurement.\footnote{On the other hand,
in ED --- just as in the Copenhagen interpretation --- other observables such
as energy or momentum do not in general have definite values; their values are
created by the act of measurement. These other quantities are epistemic in
that they do not reflect properties of the particles but of the wave
function.} In ED positions are in general \emph{unknown}; they are the
quantities to be inferred.

The position of each particle will be denoted by $x_{n}^{a}$ where the index
$n$ $=1\ldots N$ labels the particle and $a=1,2,3$ its spatial coordinates.
The position of the system in configuration space $\mathbf{X}_{N}%
=\mathbf{X}\times\ldots\times\mathbf{X}$ will be denoted either by $x$ or by
the components $x^{A}$ where $A=(n,a)$, and the corresponding volume element
is $d^{3N}x=dx$.

The second assumption is that in addition to the particles there also exist
some other variables denoted $y$ \cite{Caticha 2010a}\cite{Caticha 2017}. This
assumption is not unreasonable: the world does contain stuff beyond the $N$
particles of interest. It is also most fortunate that we need not be too
specific about these $y$ variables. It turns out that the relevant information
is conveyed by their entropy,%

\begin{equation}
S(x)=-\int dy\,p(y|x)\log\frac{p(y|x)}{q(y)}~, \label{entropy a}%
\end{equation}
where we assume that the probability distribution $p(y|x)$ depends on the
location $x$ of the particles and $q(y)$ is some unspecified underlying measure.

Having identified the microstates $(x,y)\in\mathbf{X}_{N}\times\mathbf{Y}$ we
tackle the dynamics. The goal is to find the probability density $P(x^{\prime
}|x)$ for the transition from an initial $x$ to a new $x^{\prime}$. Since both
$x^{\prime}$ and the corresponding $y^{\prime}$ are unknown the relevant space
is not just $\mathbf{X}_{N}$ but $\mathbf{X}_{N}\times\mathbf{Y}$. The
distribution we seek is the joint distribution $P(x^{\prime},y^{\prime}|x,y)$.
It is found by maximizing the appropriate entropy,
\begin{equation}
\mathcal{S}[P,Q]=-\int dx^{\prime}dy^{\prime}\,P(x^{\prime},y^{\prime
}|x,y)\log\frac{P(x^{\prime},y^{\prime}|x,y)}{Q(x^{\prime},y^{\prime}|x,y)}~,
\label{entropy joint a}%
\end{equation}
relative to a joint prior $Q(x^{\prime},y^{\prime}|x,y)$ and subject to the
appropriate constraints.

\textbf{The prior--} We adopt a prior $Q(x^{\prime},y^{\prime}|x,y)$ that
represents a state of extreme ignorance: knowledge of $x^{\prime}$ tells us
nothing about $y^{\prime}$ and vice versa. This is a product, $Q(x^{\prime
},y^{\prime}|x,y)=Q(x^{\prime}|x,y)Q(y^{\prime}|x,y)$, in which $Q(x^{\prime
}|x,y)dx^{\prime}$ and $Q(y^{\prime}|x,y)dy^{\prime}$ are
uniform,\footnote{Strictly uniform non-normalizable priors can be
mathematically problematic but here no such difficulties arise. By
\textquotedblleft uniform\textquotedblright\ we actually mean any distribution
that is essentially flat over the support of the posterior which in our case
will be infinitesimally narrow.} that is, proportional to the respective
volume elements, $d^{3N}x=dx$ and $q(y)dy$. Since proportionality constants
have no effect on the entropy maximization, the joint prior is
\begin{equation}
Q(x^{\prime},y^{\prime}|x,y)=q(y^{\prime})~. \label{prior}%
\end{equation}

\textbf{The constraints--} We first write the posterior as a product,%
\begin{equation}
P(x^{\prime},y^{\prime}|x,y)=P(x^{\prime}|x,y)P(y^{\prime}|x^{\prime},x,y)\,.
\label{joint prob}%
\end{equation}
We require that the new $x^{\prime}$ depends only on $x$ so we set
$P(x^{\prime}|x,y)=P(x^{\prime}|x)$.\ We also require that the uncertainty in
$y^{\prime}$ depends only on $x^{\prime}$, $P(y^{\prime}|x^{\prime
},x,y)=p(y^{\prime}|x^{\prime})$. Therefore, the first constraint is
\begin{equation}
P(x^{\prime},y^{\prime}|x,y)=P(x^{\prime}|x)p(y^{\prime}|x^{\prime})~.
\label{constraint p}%
\end{equation}
To implement it substitute (\ref{prior}) and (\ref{constraint p}) into
(\ref{entropy joint a}),
\begin{equation}
\mathcal{S}[P,Q]=-\int dx^{\prime}\,P(x^{\prime}|x)\log P(x^{\prime}|x)+\int
dx^{\prime}\,P(x^{\prime}|x)S(x^{\prime})~, \label{entropy joint b}%
\end{equation}
where $S(x)$ is given in eq.(\ref{entropy a}). Next, the continuity of the
motion is enforced by requiring that the steps $\Delta x_{n}^{a}$ from
$x_{n}^{a}$ to $x_{n}^{\prime a}=x_{n}^{a}+\Delta x_{n}^{a}$ taken by each
individual particle be infinitesimally short. This is implemented by imposing
$N$ independent constraints,
\begin{equation}
\int dx^{\prime}\,P(x^{\prime}|x)\Delta x_{n}^{a}\Delta x_{n}^{b}\delta
_{ab}=\langle\Delta x_{n}^{a}\Delta x_{n}^{b}\rangle\delta_{ab}=\kappa
_{n}~,\qquad(n=1\ldots N)\,. \label{constraint kappa n}%
\end{equation}
where repeated indices are summed over and we eventually take the limit
$\kappa_{n}\rightarrow0$. The $\kappa_{n}$'s are chosen to be constant to
reflect the translational symmetry of the space $\mathbf{X}$ and they are
$n$-dependent in order to accommodate non-identical particles.

\textbf{The transition probability--} Varying $P(x^{\prime}|x)$ to maximize
(\ref{entropy joint b})\ subject to (\ref{constraint kappa n}) and
normalization gives
\begin{equation}
P(x^{\prime}|x)=\frac{1}{\zeta}\exp\left[  S(x^{\prime})-\frac{1}{2}%
{\textstyle\sum\nolimits_{n}}
\alpha_{n}\delta_{ab}\Delta x_{n}^{a}\Delta x_{n}^{b}\right]  ~,
\label{trans prob a}%
\end{equation}
where $\zeta$ is a normalization constant and the Lagrange multipliers
$\alpha_{n}$ are chosen to implement the constraints
eq.(\ref{constraint kappa n}). In eq.(\ref{trans prob a}) it is clear that the
infinitesimally short steps are obtained in the limit of large $\alpha_{n}$.
It is therefore useful to Taylor expand,
\begin{equation}
S(x^{\prime})=S(x)+%
{\textstyle\sum\nolimits_{n}}
\Delta x_{n}^{a}\frac{\partial S}{\partial x_{n}^{a}}+\ldots
\end{equation}
and rewrite $P(x^{\prime}|x)$ as
\begin{equation}
P(x^{\prime}|x)=\frac{1}{Z}\exp\left[  -\frac{1}{2}%
{\textstyle\sum\nolimits_{n}}
\alpha_{n}\,\delta_{ab}\left(  \Delta x_{n}^{a}-\langle\Delta x_{n}^{a}%
\rangle\right)  \left(  \Delta x_{n}^{b}-\langle\Delta x_{n}^{b}%
\rangle\right)  \right]  ~, \label{trans prob b}%
\end{equation}
where $Z$ is a new normalization constant and $\Delta x_{n}^{a}$ is given by
eq.(\ref{Delta x}) below.

To find how these short steps accumulate we introduce time as a book-keeping
device. As discussed in \cite{Caticha 2010a}-\cite{Caticha 2015} entropic time
is measured by the fluctuations themselves (see eq.(\ref{fluc}) below) which
leads to the choice
\begin{equation}
\alpha_{n}=\frac{m_{n}}{\eta\Delta t}~, \label{alpha n}%
\end{equation}
where $\Delta t$ is the time taken by the short step, the $m_{n}$ are
particle-specific constants that will be called \textquotedblleft
masses\textquotedblright,\ and $\eta$ is a constant that fixes the units of
time relative to those of length and mass. A generic displacement is then
expressed as an expected drift plus a fluctuation,
\begin{equation}
\Delta x_{n}^{a}=\Delta x^{A}=b^{A}\Delta t+\Delta w^{A}~, \label{Delta x}%
\end{equation}
where $b^{A}(x)$ is the drift velocity,
\begin{equation}
\langle\Delta x^{A}\rangle=b^{A}\Delta t\quad\text{with}\quad b^{A}=\frac
{\eta}{m_{n}}\delta^{AB}\partial_{B}S=\eta m^{AB}\partial_{B}S~,
\label{drift velocity}%
\end{equation}
and $\partial_{A}=\partial/\partial x_{n}^{a}$; $m_{AB}=m_{n}\delta_{AB}$ is
the \textquotedblleft mass\textquotedblright\ tensor and $m^{AB}=\delta
^{AB}/m_{n}$ is its inverse. The fluctuations $\Delta w^{A}$ satisfy,
\begin{equation}
\langle\Delta w^{A}\rangle=0\quad\text{and}\quad\langle\Delta w^{A}\Delta
w^{B}\rangle=\frac{\eta}{m_{n}}\delta^{AB}\Delta t=\eta m^{AB}\Delta t~.
\label{fluc}%
\end{equation}
Thus ED leads to the non-differentiable trajectories that are characteristic
of a Brownian motion.

\textbf{The Fokker-Planck equation--} Once the probability for a single short
step is found, eq.(\ref{trans prob b}), the accumulation of many short steps
leads to a probability distribution $\rho(x,t)$ in configuration space that
obeys a Fokker-Planck equation (FP), \cite{Caticha 2012}\cite{Caticha
2010a}\cite{Caticha 2014a}
\begin{equation}
\partial_{t}\rho=-%
{\textstyle\sum\nolimits_{n}}
\partial_{na}\left(  \rho v_{n}^{a}\right)  =-\partial_{A}\left(  \rho
v^{A}\right)  ~, \label{FP b}%
\end{equation}
where $v^{A}$ is the velocity of the probability flow in configuration space
or \emph{current velocity}. It is given by
\begin{equation}
v^{A}=m^{AB}\partial_{B}\Phi_{0}\quad\text{and}\quad\Phi_{0}=\eta S-\eta
\log\rho^{1/2} \label{curr}%
\end{equation}
where $\Phi_{0}$ will be called the \emph{phase}.

\textbf{Hamiltonian entropic dynamics--} The FP eq.(\ref{FP b}) describes a
\emph{standard diffusion} of a single dynamical field, $\rho(x)$, that evolves
in response to a non-dynamical field\ given by the entropy $S(x)$. In
contrast, a \emph{quantum dynamics} includes a second dynamical field, the
phase of the wave function. In ED this evolving phase is introduced by
continuously updating the constraint (\ref{constraint p}) which allows the
entropy $S(x)$, or equivalently the phase $\Phi_{0}(x)$, to become dynamical.

First we note that without loss of generality we can always find a functional
$\tilde{H}[\rho,\Phi_{0}]$ so that $\partial_{t}\rho=\delta\tilde{H}%
/\delta\Phi_{0}$ reproduces the FP equation (\ref{FP b}). The specific
updating rule for $S$ or $\Phi_{0}$ is inspired by an idea of Nelson's
\cite{Nelson 1979}: requiring that $\Phi_{0}$ be updated in such a way that
the functional $\tilde{H}[\rho,\Phi_{0}]$ be conserved leads to Hamilton's
equations \cite{Caticha 2014b},
\begin{equation}
\partial_{t}\rho=\frac{\delta\tilde{H}}{\delta\Phi_{0}}\quad\text{and}%
\quad\partial_{t}\Phi_{0}=-\frac{\delta\tilde{H}}{\delta\rho}~.
\label{Hamilton}%
\end{equation}
$\tilde{H}[\rho,\Phi_{0}]$ is the \textquotedblleft ensemble\textquotedblright%
\ Hamiltonian. The second equation in (\ref{Hamilton}) is a Hamilton-Jacobi
equation (HJ). Additional arguments from information geometry \cite{Caticha
2014b} can then be invoked to suggest that the natural choice of $\tilde{H}$
is%
\begin{equation}
\tilde{H}[\rho,\Phi_{0}]=\int dx\,\rho\left[  \frac{1}{2}m^{AB}\partial
_{A}\Phi_{0}\partial_{B}\Phi_{0}+V+\xi m^{AB}\frac{1}{\rho^{2}}\partial
_{A}\rho\partial_{B}\rho\right]  ~. \label{Hamiltonian}%
\end{equation}
The first term in the integrand is the \textquotedblleft
kinetic\textquotedblright\ term that reproduces the FP equation (\ref{FP b}).
The second term represents the simplest non-trivial interaction and introduces
the standard potential $V(x)$. The third term, motivated by information
geometry, is the trace of the Fisher information and is called the
\textquotedblleft quantum\textquotedblright\ potential. The parameter $\xi
$\ controls the relative contributions of the two potentials: $\xi=0$ leads to
a stochastic classical mechanics; $\xi>0$ leads to quantum theory --- in fact,
$\xi$ \emph{defines} Planck's constant as $\hbar=(8\xi)^{1/2}$.

\textbf{The Schr\"{o}dinger equation--} To conclude this brief review of ED we
note that at this point the dynamics is fully specified by equations
(\ref{Hamilton}) and (\ref{Hamiltonian}). We can combine $\rho$ and $\Phi_{0}$
into a single complex function, $\Psi_{0}=\rho^{1/2}\exp(i\Phi_{0}/\hbar)$.
Then the pair of Hamilton's equations (\ref{Hamilton}) can be rewritten as a
single complex Schr\"{o}dinger equation that is explicitly linear,%
\begin{equation}
i\hbar\partial_{t}\Psi_{0}=-\frac{\hbar^{2}}{2}m^{AB}\partial_{A}\partial
_{B}\Psi_{0}+V\Psi_{0}~. \label{sch c}%
\end{equation}
However, even though eqs.(\ref{Hamilton}) can be written in the form
(\ref{sch c}), this does not mean that they are equivalent to the full quantum
theory. The problem is that eqs.(\ref{Hamilton}) only reproduce a subset of
all the wave functions required by quantum mechanics. More specifically since
both $S(x)$ and $\rho(x)$ are single-valued --- the total change as one moves
in a closed path vanishes,
\begin{equation}
\Delta S=%
{\displaystyle\oint_{\Gamma}}
d\ell^{A}\partial_{A}S=0\quad\text{and}\quad\Delta\rho=%
{\displaystyle\oint_{\Gamma}}
d\ell^{A}\partial_{A}\rho=0~,
\end{equation}
so that both $\Phi_{0}$ and $\Psi_{0}$ are single-valued too. The
single-valuedness of $\Psi_{0}$ is precisely what we want, but the
single-valuedness of $\Phi_{0}$ is too restrictive. It excludes, for example,
eigenstates of angular momentum that have manifestly multi-valued phases
($\Psi\propto e^{im\phi}$ where $\phi$ is the azimuthal angle and $m$ is an integer).

\section{Gauge symmetry and multi-valued phases}

A minimal ED was derived in the previous section. A richer dynamics that
allows additional interactions can be achieved by imposing additional constraints.

\textbf{Additional constraints--} We assume that the motion of each particle
is affected by an additional potential field $\varphi(x)$ where $x\in
\mathbf{X}$ is a point in 3D space with the topological properties of an angle
($\varphi(x)$ and $\varphi(x)+2\pi$ describe the same angle). We further
assume that these angles can be redefined by different amounts $\chi(x)$ at
different places, that is, the origin from which these angles are measured can
be set independently at each $x$. This is a local gauge symmetry and it
immediately raises the question of how can one compare angles at different
locations in order to define derivatives. The answer is well known: introduce
a \emph{connection} field, a vector potential $A_{a}(x)$ that defines which
angle at $x+\Delta x$ is the \textquotedblleft same\textquotedblright\ as the
angle at $x$. This is implemented by imposing that as $\varphi\rightarrow
\varphi+\chi$ then the connection transforms as $A_{a}\rightarrow
A_{a}+\partial_{a}\chi$ so that the corrected derivative $\partial_{a}%
\varphi-A_{a}$ remains invariant.\footnote{Note that since $\varphi$ is
dimensionless the vector potential $A_{a}$ has units of inverse length and
this implicitly defines the units of electric charge. These are not the units
conventionally adopted in electromagnetism.}

To derive an ED that incorporates interactions gauge invariant interactions
with these potentials, in addition to (\ref{constraint kappa n}) and
normalization, for each particle we impose the constraint
\begin{equation}
\langle\Delta x^{a}\rangle\left[  \partial_{a}\varphi(x_{n})-A_{a}%
(x_{n})\right]  =\kappa_{n}^{\prime}(x_{n})~,\quad(n=1\ldots N)
\label{constraint phi}%
\end{equation}
where $\kappa_{n}^{\prime}(x_{n})$ are functions to be specified below.

The transition probability $P(x^{\prime}|x)$ that maximizes the entropy
$\mathcal{S}[P,Q]$ in (\ref{entropy joint b}) is
\begin{equation}
P(x^{\prime}|x)=\frac{1}{\zeta}\exp\left[  S(x^{\prime})-%
{\textstyle\sum\nolimits_{n}}
\left(  \frac{\alpha_{n}}{2}\delta_{ab}\Delta x_{n}^{a}\Delta x_{n}^{b}%
-\beta_{n}\left(  \partial_{na}\varphi(x_{n})-A_{a}(x_{n})\right)  \Delta
x_{n}^{a}\right)  \right]
\end{equation}
where $\partial_{na}=\partial/\partial x_{n}^{a}$, $\alpha_{n}$ and $\beta
_{n}$ are Lagrange multipliers, and $\zeta$ is a normalization constant. For
large $\alpha_{n}$ Taylor expand $S(x^{\prime})$ about $x$, and use
eq.(\ref{alpha n}), then, as in eq.(\ref{Delta x}) a generic displacement
$\Delta x^{A}$ can be expressed in terms of a expected drift plus a
fluctuation, $\Delta x^{A}=b^{A}\Delta t+\Delta w^{A}$, but the drift velocity
(\ref{drift velocity}) now includes a new term,%
\begin{equation}
b_{n}^{a}=\frac{\eta}{m_{n}}\delta^{ab}\left[  \partial_{nb}\{S(x)+\beta
_{n}\varphi(x_{n})\}-\beta_{n}A_{b}(x_{n})\right]  ~,
\end{equation}
while the fluctuations $\Delta w^{A}$, eq.(\ref{fluc}), remain unchanged.

\textbf{Hamilton's equations --} As before, the accumulation of many short
steps leads to the FP equation (\ref{FP b}), but now the current velocity
$v^{A}=v_{n}^{a}$ must be suitably modified,
\begin{equation}
v^{A}=m^{AB}\left(  \partial_{B}\Phi-\bar{A}_{B}\right)  \quad\text{with}%
\quad\Phi=\eta(S+\bar{\varphi}-\log\rho^{1/2})~,\label{curr b}%
\end{equation}
where we introduced the configuration space quantities,
\begin{equation}
\bar{A}_{A}(x)=\eta\beta_{n}A_{a}(x_{n})\quad\text{and}\quad\bar{\varphi}(x)=%
{\textstyle\sum\nolimits_{n}}
\beta_{n}\varphi(x_{n})~.
\end{equation}
where $A=(n,a)$. Note that $v^{A}$ is gauge invariant. The new ensemble
Hamiltonian $\tilde{H}$, eq.(\ref{Hamiltonian}), is%
\begin{equation}
\tilde{H}[\rho,\Phi]=\int dx\,\left[  \frac{1}{2}\rho m^{AB}\left(
\partial_{A}\Phi-\bar{A}_{A}\right)  \left(  \partial_{B}\Phi-\bar{A}%
_{B}\right)  +\rho V+\,\frac{\hbar^{2}}{8\rho}m^{AB}\partial_{A}\rho
\partial_{B}\rho\right]  ~,\label{Hamiltonian b}%
\end{equation}
and the new FP equation now reads,
\begin{equation}
\partial_{t}\rho=-\partial_{A}\left[  \rho m^{AB}\left(  \partial_{B}\Phi
-\bar{A}_{B}\right)  \right]  =\frac{\delta\tilde{H}}{\delta\Phi}~.
\end{equation}
As before, the requirement that $\tilde{H}$ be conserved for arbitrary initial
conditions amounts to imposing the conjugate Hamilton equation,
eq.(\ref{Hamilton}), which leads to the Hamilton-Jacobi equation,
\begin{equation}
\partial_{t}\Phi=-\frac{\delta\tilde{H}}{\delta\rho}=-\frac{1}{2}m^{AB}\left(
\partial_{A}\Phi-\bar{A}_{A}\right)  \left(  \partial_{B}\Phi-\bar{A}%
_{B}\right)  -V+\frac{\hbar^{2}}{2}m^{AB}\frac{\partial_{A}\partial_{B}%
\rho^{1/2}}{\rho^{1/2}}~~.\label{HJ d}%
\end{equation}
Finally, we combine $\rho$ and $\Phi$ into a single wave function, $\Psi
=\rho^{1/2}\exp(i\Phi/\hbar)$, to obtain the linear Schr\"{o}dinger equation,
\begin{equation}
i\hbar\partial_{t}\Psi=-%
{\displaystyle\sum\nolimits_{n}}
\frac{\hbar^{2}}{2m_{n}}\delta^{ab}(\frac{\partial}{\partial x_{n}^{a}}%
-\frac{i}{\hbar}\eta\beta_{n}A_{a}(x_{n}))(\frac{\partial}{\partial x_{n}^{b}%
}-\frac{i}{\hbar}\eta\beta_{n}A_{b}(x_{n}))\Psi+V\Psi~.\label{sch d}%
\end{equation}

\section{Discussion}

\textbf{Electric charges are Lagrange multipliers-- }Recalling the standard
expression for covariant derivatives,
\begin{equation}
\frac{\partial}{\partial x_{n}^{a}}-\frac{iq_{n}}{\hbar c}A_{a}(x_{n})~,
\end{equation}
($q_{n}$ is the electric charge of particle $n$ and $c$ is the speed of light)
shows that (\ref{sch d}) is indeed the Schr\"{o}dinger equation provided the
multipliers $\beta_{n}$ are chosen to be particle-dependent \emph{constants}
that are related to electric charges by
\begin{equation}
\beta_{n}=\frac{q_{n}}{\eta c}\quad\text{or}\quad q_{n}=c\eta\beta_{n}~.
\label{charge a}%
\end{equation}
Thus, in ED \emph{electric charges are Lagrange multipliers} that measure the
strength of the particles' coupling to the $\varphi_{n}$ and $A_{a}$ potentials.

\textbf{Single-valued wave functions, quantized circulation, and quantized
charges-- }The success of any framework for inference such as ED depends on
identifying the correct constraints. The choice of constraints in section 2
succeeds in reproducing many of the features of quantum theory including a
linear Schr\"{o}dinger equation but is ultimately unsatisfactory because it
leads to single-valued wave functions with single-valued phases that fail to
include all quantum states.

The choice of constraints adopted in section 3 represent an improvement
because they take into account the relation between quantum phases and gauge
symmetry. However, the wave functions $\Psi$ obtained for generic choices of
the multipliers $\beta_{n}$ are also problematic in that they give
multi-valued phases $\Phi$, eq.(\ref{curr b}), that lead to multi-valued wave
functions. Indeed, since $\varphi$ is an angle the integral over a closed loop
$\Gamma_{n}$ in which all particles except $n$ are kept fixed gives
\begin{equation}
\Delta\varphi=%
{\displaystyle\oint_{\Gamma_{n}}}
d\ell_{n}^{a}\partial_{na}\varphi=2\pi\nu(\Gamma_{n})~,
\end{equation}
where $\nu(\Gamma_{n})$ is an integer that depends on the loop $\Gamma_{n}$.
Since $S$ and $\log\rho$ are single-valued, from (\ref{curr b}), we have
\begin{equation}
\Delta\frac{\Phi}{\hbar}=%
{\displaystyle\oint_{\Gamma_{n}}}
d\ell_{n}^{a}\partial_{na}\frac{\Phi}{\hbar}=\frac{\eta\beta_{n}}{\hbar}%
{\displaystyle\oint_{\Gamma_{n}}}
d\ell_{n}^{a}\partial_{na}\varphi=\frac{\eta\beta_{n}}{\hbar}2\pi\nu
(\Gamma_{n})~, \label{circ b}%
\end{equation}
so that $\Psi$ is not single-valued.

Unfortunately, this means that even though (\ref{sch d}) is linear, its
linearity is in conflict with the underlying probabilistic structure. To see
the problem consider two multivalued ED solutions of (\ref{sch d}), $\Psi_{1}$
and $\Psi_{2}$. Their magnitudes $|\Psi_{1}|^{2}=\rho_{1}$ and $|\Psi_{2}%
|^{2}=\rho_{2}$ are single-valued because they are probability densities.
However, even though $\alpha_{1}\Psi_{1}+\alpha_{2}\Psi_{2}=\Psi_{3}$ is also
a solution, it turns out that its magnitude $|\Psi_{3}|^{2}\ $will in general
turn out to be multivalued which precludes a probabilistic interpretation
\cite{Merzbacher 1962}. Mere linearity is not enough. The condition for the
linear and probabilistic structures to be compatible with each other is that
wave functions be single-valued.

Inspection of Eq.(\ref{circ b}) for arbitrary loops shows that the choice of
constraint (\ref{constraint phi}) --- that is, the choice of $\beta_{n}$ ---
that leads to single-valued wave functions is
\begin{equation}
\frac{\eta\beta_{n}}{\hbar}=\mu\label{alpha prime}%
\end{equation}
where $\mu$ is an integer.

Equation (\ref{charge a}) then shows that electric charges must be quantized
in units of a basic charge $q=\hbar c$,
\begin{equation}
\frac{\eta\beta_{n}}{\hbar}=\frac{q_{n}}{\hbar c}=\mu\quad\text{or}\quad
q_{n}=\mu q~.
\end{equation}
Changing to conventional units for charges and potentials is straightforward;
just rescale $\lambda q_{n}=q_{n}^{\prime}$ and $A_{a}/\lambda=A_{a}^{\prime}$
so that $q_{n}A_{a}=q_{n}^{\prime}A_{a}^{\prime}$.

\textbf{Conclusion-- }The equivalence of ED and quantum mechanics with wave
functions that remain single-valued even for multi-valued phases is achieved
by imposing constraints that recognize the intimate relation between quantum
phases and gauge symmetry. The condition for compatibility between the
probabilistic and linear structures is that charges be quantized.

\paragraph{Acknowledgments}

We would like to thank M. Abedi, D. Bartolomeo, C. Cafaro, N. Caticha, S.
DiFranzo, A. Giffin, S. Ipek, D.T. Johnson, K. Knuth, S. Nawaz, M. Reginatto,
C. Rodr\'{\i}guez, K. Vanslette, for many discussions on entropy, inference
and quantum mechanics.

\end{document}